\documentclass[12pt]{article}
\usepackage[pctex32]{graphicx}
\linethickness{0.4pt}

\begin{document}
\vskip 4.0 true cm
\begin{center}
{\LARGE \bf Long-range Effect on the Curie Temperature of Ferroelectric Films}
\vskip 0.8 true cm
{{\sc Dong-Lai Yao} $(a)$, {\sc Yin-Zhong Wu} ${(a,b)}$}
\vskip 0.2 true cm
{\it a. Department of Physics, Suzhou University, Suzhou, 215006, China}\\
{\it b. Department of Physics, Changshu College, Changshu, 215500, China}\\
\end{center}
\vspace{0.5cm}
\begin{center}
{{\sc Zhen-Ya Li}}\\
{\it CCAST(World Laboratory), P.O.Box 8730, Beijing 100080, China}\\
{\it and Department of Physics, Suzhou University, Suzhou, 215006, China$^*$}\\
\vspace{2.0cm}
\end{center}

\begin{abstract}
In this paper, the Curie temperature of ferroelectric films is studied using spin-1/2 transverse Ising model with long-range interaction within the framework of the effective-field theory. The dependence of the Curie temperature on the thickness of the film, the surface interaction and the transverse field were investigated. It is assumed that the long-range interaction decays with the distance between the pseudo-spins as a power law. The dependence of the Curie temperature and the critical transverse field on the long-range exponent are obtained.
\end{abstract}

\vspace{1.0cm}



*mailing address in China

E-mail: zyli@suda.edu.cn

\newpage
{\bf I. Introduction}
\quad

The pseudo-spin theory based on the Ising model with a transverse field (TIM) was first introduced by de Gennes [1] to describe the phase transition in order-disorder type ferroelectrics [2]. Under this model, it is assumed that a ferroelectric is composed of pseudo-spins with interactions, and this model had been applied to other systems successfully [3,4]. In most previous works [5-8], only the nearest neighbor interaction was considered. In ferromagnets where short-range interactions dominated, this simplification seems reasonable. But in ferroelectrics, where long-range interactions are thought to be dominant [9], the interactions between pseudo-spins over long distance should be taken into consideration. In previous works, the long-range coupling interaction has been considered [10-12], it is found that the long-range interaction plays a vital role in ferroelectric structure. Theoretically, there are two methods to investigate properties of ferroelectrics, Ginzburg-Landau theory and the microscopic transverse Ising model. The long-range interaction in ferroelectric film has been studied by using TIM and the mean-field theory (MFT). However, the correlation between some spins is neglected in MFT. In order to study the critical behavior of ferroelectric film, some works have used TIM model within the frame work of the effective-field theory (EFT), however the long-range interaction was not considered[5-6]. It needs to use an approach, which is superior to the MFT. In this paper, we study the phase transition properties of ferroelectric film by use of EFT with correlation within the framework of TIM model, and the long-range interaction is considered in the model.

The effective-field theory, which is based on the Ising spin identities and the differential operator technique, theoretically, is superior to the mean-field theory. The EFT has been used for the investigation of phase diagrams [5], and successfully applied to a variety of physical systems, such as magnetic thin films and superlattice [12,13], ferroelectric films [14] and ferroelectric superlattices [15]. In this paper, it is supposed that the interaction between the pseudo-spins decays as $1/r^\delta$ [16], where $\delta$ represents the decaying exponent. The case of $\delta\rightarrow\infty$ corresponds to the short-range interaction case, and the case of $\delta\rightarrow0$ reduces to the infinite-range coupling one. According to our numerical results, we find that the long-range interaction has great influence on the phase transition temperature and the critical transverse field $({\it \Omega}_c)$. They both increase as the long-range interaction becomes stronger.

\vspace{0.5cm}
{\bf II. Theory and Formulation}

We consider a film consists of N layers. Each layer is defined on the x-y plane and with pseudo-spin sites on a square lattice. As shown in Fig. 1, the 1-st and the N-th layers are the surface layers, in which the interaction constant is  different from that in the inner layers. $J_s$ denotes the interaction constant of the pseudo-spins at the surface layer, and $J_b$ that of the inner layer. We assume, for simplicity, that the interlayer interactions is $J_b$.
In order to consider the long-range interaction in the ferroelectric thin film, we introduce the interaction constant $J=J_{i,j}/r_{i,j}^\delta$ between pseudo-spins at i-th and j-th sites as in Ref. [16,17]. The system is described by the Ising Hamiltonian with a transverse field as

\begin{equation}
H =  - \sum\limits_{i,j} {\frac{{J_{i,j} }}{{r_{i,j}^\delta  }}S_i^z S_j^z }  - \sum\limits_i {{\it {\it \Omega}} _i S_i^x },
\end{equation}

where $S_i^z$ and $S_i^x$ are the components of the pseudo spin operator at site i, $r_{i,j}$ is the distance between the i-th and j-th pseudo-spin, $\delta$ is decaying exponent.

According to the effective-field theory with correlation ${[18,19]}$. The average value of pseudo spins in the i-th (i=1,2,3,...,N) layer can be given by

\begin{equation}
R_i  = \left\langle {S_i^z } \right\rangle  = \prod\limits_j {[\cosh (\frac{{J_{i,j} }}{{r_{i,j}^\delta  }}\nabla ) + R_{j} \\\sinh (\frac{{J_{i,j} }}{{r_{i,j}^\delta  }}\nabla )]\left. {F(x)} \right|_{x = 0} }, 
\end{equation}

$\Pi$ runs over all of the neighbors of site i, $\nabla=\frac{\partial}{\partial x}$  is the differential operator. The function $F(x)$ is given by $F(x)=\frac{x}{y}tanh(\beta y)$, where $\beta=\frac{1}{k_B T}$, $y=\sqrt{x^2+{\it \Omega}^2}$.

The phase transition temperature is an important quantity in studying the critical properties of the ferroelectric film. Near the critical point, the polarization in each layer is very small, thus we can linearize Eq. (2). This leads to the matrix equation of the form $Z_{R_i}$= 0.

For instance, we take N=10, and considering the 3-rd nearest-neighbor, then

\begin{eqnarray}
\nonumber R_1  &=& [\cosh (J_s \nabla ) + R_1 \sinh (J_s \nabla )]^4 [\cosh (J_b \nabla ) + R_2 \sinh (J_b \nabla )]\\
&&[\cosh (\frac{{J_s }}{{\sqrt 2 ^\delta  }}\nabla ) + R_1 \sinh (\frac{{J_s }}{{\sqrt 2 ^\delta  }}\nabla )]^4 [\cosh (\frac{{J_b }}{{\sqrt 2 ^\delta  }}\nabla ) + R_2 \sinh (\frac{{J_b }}{{\sqrt 2 ^\delta  }}\nabla )]^4\\\nonumber
&&[\cosh (\frac{{J_b }}{{\sqrt 3 ^\delta  }}\nabla ) + R_2 \sinh (\frac{{J_b }}{{\sqrt 3 ^\delta  }}\nabla )]^4 \left. {F(x)} \right|_{x = 0}, \\\nonumber
R_2  &=& [\cosh (J_s \nabla ) + R_1 \sinh (J_s \nabla )][\cosh (J_b \nabla ) + R_2 \sinh (J_b \nabla )]^4\\\nonumber
&&[\cosh (J_b \nabla ) + R_3 \sinh (J_b \nabla )][\cosh (\frac{{J_b }}{{\sqrt 2 ^\delta  }}\nabla ) + R_1 \sinh (\frac{{J_b }}{{\sqrt 2 ^\delta  }}\nabla )]^4 \\
&&[\cosh (\frac{{J_b }}{{\sqrt 2 ^\delta  }}\nabla ) + R_2 \sinh (\frac{{J_b }}{{\sqrt 2 ^\delta  }}\nabla )]^4 [\cosh (\frac{{J_b }}{{\sqrt 2 ^\delta  }}\nabla ) + R_3 \sinh (\frac{{J_b }}{{\sqrt 2 ^\delta  }}\nabla )]^4\\\nonumber 
&&[\cosh (\frac{{J_b }}{{\sqrt 3 ^\delta  }}\nabla ) + R_1 \sinh (\frac{{J_b }}{{\sqrt 3 ^\delta  }}\nabla )]^4 [\cosh (\frac{{J_b }}{{\sqrt 3 ^\delta  }}\nabla ) + R_3 \sinh (\frac{{J_b }}{{\sqrt 3 ^\delta  }}\nabla )]^4 \left. {F(x)} \right|_{x = 0}, \\\nonumber
&.&\\\nonumber
&.&\\\nonumber
&.&\\\nonumber
R_i&=&[\cosh (J_s \nabla ) + R_{i - 1} \sinh (J_s \nabla )][\cosh (J_b \nabla ) + R_i \sinh (J_b\nabla )]^4 \\\nonumber
&&[\cosh (J_b \nabla ) + R_{i + 1} \sinh (J_b \nabla )][\cosh (\frac{{J_b }}{{\sqrt 2 ^\delta  }}\nabla ) + R_{i - 1} \sinh (\frac{{J_b }}{{\sqrt 2 ^\delta  }}\nabla )]^4\\
&&[\cosh (\frac{{J_b }}{{\sqrt 2 ^\delta  }}\nabla ) + R_i \sinh (\frac{{J_b }}{{\sqrt 2 ^\delta  }}\nabla )]^4[\cosh (\frac{{J_b }}{{\sqrt 2 ^\delta  }}\nabla ) + R_{i + 1} \sinh (\frac{{J_b }}{{\sqrt 2 ^\delta  }}\nabla )]^4 \\\nonumber
&&[\cosh (\frac{{J_b }}{{\sqrt 3 ^\delta  }}\nabla ) + R_{i - 1} \sinh (\frac{{J_b }}{{\sqrt 3 ^\delta  }}\nabla )]^4 [\cosh (\frac{{J_b }}{{\sqrt 3 ^\delta  }}\nabla ) + R_{i + 1} \sinh (\frac{{J_b }}{{\sqrt 3 ^\delta  }}\nabla )]^4 \left. {F(x)} \right|_{x = 0},\\\nonumber
&.&\\\nonumber
&.&\\\nonumber
&.&\\\nonumber
R_9  &=& [\cosh (J_s \nabla ) + R_{10} \sinh (J_s \nabla )][\cosh (J_b \nabla ) + R_9 \sinh (J_b \nabla )]^4 \\\nonumber
&&[\cosh (J_b \nabla ) + R_8 \sinh (J_b \nabla )][\cosh (\frac{{J_b }}{{\sqrt 2 ^\delta  }}\nabla ) + R_{10} \sinh (\frac{{J_b }}{{\sqrt 2 ^\delta  }}\nabla )]^4\\
&& [\cosh (\frac{{J_b }}{{\sqrt 2 ^\delta  }}\nabla ) + R_9 \sinh (\frac{{J_b }}{{\sqrt 2 ^\delta  }}\nabla )]^4 [\cosh (\frac{{J_b }}{{\sqrt 2 ^\delta  }}\nabla ) + R_8 \sinh (\frac{{J_b }}{{\sqrt 2 ^\delta  }}\nabla )]^4\\\nonumber
&&[\cosh (\frac{{J_b }}{{\sqrt 3 ^\delta  }}\nabla ) + R_{10} \sinh (\frac{{J_b }}{{\sqrt 3 ^\delta  }}\nabla )]^4 [\cosh (\frac{{J_b }}{{\sqrt 3 ^\delta  }}\nabla ) + R_8 \sinh (\frac{{J_b }}{{\sqrt 3 ^\delta  }}\nabla )]^4 \left. {F(x)} \right|_{x = 0}, \\\nonumber
R_{10}&=&[\cosh (J_s \nabla ) + R_{10} \sinh (J_s \nabla )]^4 [\cosh (J_b \nabla ) + R_9 \sinh (J_b \nabla )]\\
&&[\cosh (\frac{{J_s }}{{\sqrt 2 ^\delta  }}\nabla ) + R_{10} \sinh (\frac{{J_s }}{{\sqrt 2 ^\delta  }}\nabla )]^4 [\cosh (\frac{{J_b }}{{\sqrt 2 ^\delta  }}\nabla ) + R_9 \sinh (\frac{{J_b }}{{\sqrt 2 ^\delta  }}\nabla )]^4\\\nonumber
&&[\cosh (\frac{{J_b }}{{\sqrt 3 ^\delta  }}\nabla ) + R_9 \sinh (\frac{{J_b }}{{\sqrt 3 ^\delta  }}\nabla )]^4 \left. {F(x)} \right|_{x = 0}. \\\nonumber
\end{eqnarray}

Where $R_i$ expresses the longitudinal polarization of a pseudo-spin in the i-th layer and the i takes the value from 1 to N.

By linearizing the equations above, we obtain the matrix $Z_{R_i}=0$:

\begin{equation}
\left( {\begin{array}{*{20}c}
   {Z_{11} } & {Z_{12} } & 0 & 0 & 0 & 0 & 0 & 0 & 0 & 0  \\
   {Z_{21} } & {Z_{22} } & {Z_{21} } & 0 & 0 & 0 & 0 & 0 & 0 & 0  \\
   0 & {Z_{21} } & {Z_{22} } & {Z_{21} } & 0 & 0 & 0 & 0 & 0 & 0  \\
   0 & 0 & {Z_{21} } & {Z_{22} } & {Z_{21} } & 0 & 0 & 0 & 0 & 0  \\
   0 & 0 & 0 & {Z_{21} } & {Z_{22} } & {Z_{21} } & 0 & 0 & 0 & 0  \\
   0 & 0 & 0 & 0 & {Z_{21} } & {Z_{22} } & {Z_{21} } & 0 & 0 & 0  \\
   0 & 0 & 0 & 0 & 0 & {Z_{21} } & {Z_{22} } & {Z_{21} } & 0 & 0  \\
   0 & 0 & 0 & 0 & 0 & 0 & {Z_{21} } & {Z_{22} } & {Z_{21} } & 0  \\
   0 & 0 & 0 & 0 & 0 & 0 & 0 & {Z_{21} } & {Z_{22} } & {Z_{21} }  \\
   0 & 0 & 0 & 0 & 0 & 0 & 0 & 0 & {Z_{12} } & {Z_{11} }  \\
\end{array}} \right)= 0,
\end{equation}
where
\begin{eqnarray}
\nonumber Z_{11}  &=& [4\cosh(J_b \nabla )\cosh^4 (\frac{{J_b }}{{\sqrt 2 ^\delta  }}\nabla )\cosh^4 (\frac{{J_s }}{{\sqrt 2 ^\delta  }}\nabla )\cosh^4 (\frac{{J_b }}{{\sqrt 3 ^\delta  }}\nabla )\sinh^3 (J_s \nabla ) \\\nonumber
&&+ 4\cosh(J_b \nabla )\cosh^4 (J_s \nabla )\cosh^4 (\frac{{J_b }}{{\sqrt 2 ^\delta  }}\nabla )\cosh^4 (\frac{{J_b }}{{\sqrt 3 ^\delta  }}\nabla )\sinh^3 (\frac{{J_s }}{{\sqrt 2 ^\delta  }}\nabla )]F(x)|_{x = 0},\\
\nonumber Z_{12}  &=& [4\cosh^4 (J_s \nabla )\cosh^4 (\frac{{J_b }}{{\sqrt 2 ^\delta  }}\nabla )\cosh^4 (\frac{{J_s }}{{\sqrt 2 ^\delta  }}\nabla )\cosh^4 (\frac{{J_b }}{{\sqrt 3 ^\delta  }}\nabla )\sinh(J_b \nabla ) \\\nonumber 
&&+4\cosh(J_b \nabla )\cosh^4 (J_s \nabla )\cosh^4 (\frac{{J_s }}{{\sqrt 2 ^\delta  }}\nabla )\cosh^4 (\frac{{J_b }}{{\sqrt 3 ^\delta  }}\nabla )\sinh^3 (\frac{{J_b }}{{\sqrt 2 ^\delta  }}\nabla ) \\\nonumber 
&&+ 4\cosh(J_b \nabla )\cosh^4 (J_s \nabla )\cosh^4 (\frac{{J_b }}{{\sqrt 2 ^\delta  }}\nabla )\cosh^4 (\frac{{J_s }}{{\sqrt 2 ^\delta  }}\nabla )\sinh^3 (\frac{{J_b }}{{\sqrt 3 ^\delta  }}\nabla )]F(x)|_{x = 0},\\
\nonumber Z_{21}  &=& [\cosh^5 (J_b \nabla )\cosh^{12} (\frac{{J_b }}{{\sqrt 2 ^\delta  }}\nabla )\cosh^8 (\frac{{J_b }}{{\sqrt 3 ^\delta  }}\nabla )\sinh(J_b \nabla ) \\\nonumber
&&+ 4\cosh^6 (J_b \nabla )\cosh^8 (\frac{{J_b }}{{\sqrt 2 ^\delta  }}\nabla )\cosh^8 (\frac{{J_b }}{{\sqrt 3 ^\delta  }}\nabla )\sinh^3 (\frac{{J_b }}{{\sqrt 2 ^\delta  }}\nabla ) \\\nonumber
&&+ 4\cosh^6 (J_b \nabla )\cosh^{12} (\frac{{J_b }}{{\sqrt 2 ^\delta  }}\nabla )\cosh^4 (\frac{{J_s }}{{\sqrt 3 ^\delta  }}\nabla )\sinh^3 (\frac{{J_b }}{{\sqrt 3 ^\delta  }}\nabla )]F(x)|_{x = 0},\\
\nonumber Z_{22}  &=& [4\cosh^2 (J_b \nabla )\cosh^{12} (\frac{{J_b }}{{\sqrt 2 ^\delta  }}\nabla )\cosh^8 (\frac{{J_b }}{{\sqrt 3 ^\delta  }}\nabla )\sinh^3 (J_b \nabla ) \\\nonumber
&&+ 4\cosh^6 (J_b \nabla )\cosh^8 (\frac{{J_b }}{{\sqrt 2 ^\delta  }}\nabla )\cosh^8 (\frac{{J_s }}{{\sqrt 3 ^\delta  }}\nabla )\sinh^3 (\frac{{J_b }}{{\sqrt 2 ^\delta  }}\nabla )]F(x)|_{x = 0}.
\end{eqnarray}

We can obtain the phase transition temperature of the ferroelectric film by taking the highest temperature from the solutions of the equation $det(Z_{R_i})=0$.

\vspace{0.5cm}
{\bf III. Calculation and Discussion}

First, we putting ${\it \Omega}/J_b=0, 2.0$, $J_s/J_b=1$ into the formulation in the previous section. Fig. 2 shows the $T_c$ versus n plot for different $\delta : \delta = 1000$ (in our calculation, it is equivalent to $\delta\rightarrow\infty$ approximately), $\delta = 6.0$, and $\delta = 3.2$. In Fig. 2, the solid line and dashed line correspond to ${\it \Omega}/J_b=0$ and ${\it \Omega}/J_b=2.0$ respectively. For large $\delta (\delta = 1000)$, the long-range interaction is very weak, and it is corresponds to the short-range interaction. The smaller the value of $\delta$, the stronger the long-range coupling interaction effect. From Fig. 2, we can see the dashed line of $\delta = 1000$ recovers the curve of transition temperature of a thin ferroelectric film versus the thickness n in the Zernike (Effective-field) approximation[5]. With the increase of the thickness, the phase transition temperature of the film approaches to the bulk one.

As shown in Fig. 2, with the decrease of $\delta$, which corresponds that the long-range interaction becomes stronger, the phase transition temperature increases. This is reasonable since strong long-range interactions represent that there are more other pseudo-spins around one pseudo-spin, which have the interaction on it. As a result, it can hardly flip freely. For the multilayer ferroelectric system, strong long-range interactions make the system harder to be disordered. Consequently, the transition temperature increases.

Comparing Curve 2' in Fig. 2 in our paper with Curve 1 in Fig. 3(b) in Ref [11], we find that they are very similar in appearance for the two curves. When film thickness reaches $n=10$, the curve has almost tended to be saturate, i.e. $T_c$ tends to the bulk $T_c$. It is clear that the MFT and the EFT obtain the same results. But the difference between Curve 1' in Fig. 2 in our paper and Curve 3 in Fig. 3(b) in Ref. [11] is obvious. The two curves both represent $\delta=3.2$, but they are obtained by MFT and EFT, respectively. There is no $T_c=0$ thickness in Curve 1' in Fig. 2 in our paper, while in Curve 3 in Fig. 3(b) in Ref [11], for a range of small film thickness in which the system has been disordered at zero temperature.

It was very surprising that the Curie temperature and the polarization decrease with the increase of the long-range coupling interaction in Ref. [11]. Our result is that $T_c$ increases with the increase of the long-range interaction coupling as Fig. 2 shown. In physics, the interactions between the pseudo spins become stronger and stronger; therefore it is harder to be disordered for the system. Thus, only when the temperature is high enough, could the system become disordered. Therefore we believed that our results are more reasonable.

Comparing the solid line with the dashed line in Fig. 2, we find the $T_c$ changes with different transverse field. Given the thickness $n$ and the value of the decaying exponent $\delta$, the $T_c$ with a transverse field ${\it \Omega}/J_b=2.0$ is lower than that without transverse field (See the dashed line in Fig. 2). i.e. the transverse field makes the $T_c$ of the system lower.

The relationship between $T_c$ and the transverse field is shown in Fig. 3. We plot the $T_c$ versus ${\it \Omega}$ with different $\delta$. There exist a series of critical transverse field ${\it \Omega}_{cs}$ beyond which the system has already been disordered at $0$ K[5]. ${\it \Omega}_c$ decreases as the strength of the long-range interaction decreasing. The stronger the long-range interaction, the larger the critical transverse field ${\it \Omega}_c$.

Fig. 4 shows the Curie temperature as a function of surface coupling interaction $J_s$. We take the thickness n=10. In Fig. 4a, we take different $\delta$, which indicates the different range of the interactions. With the $J_s$ increasing, $T_c$ increases. At first, $T_c$ increases slowly. When $J_s$ reaches a certain point, $T_c$ increases linearly as the $J_s$. We also find the same result as in Fig. 2, i.e. smaller $\delta$ results in larger $T_c$. While in Fig. 4(b), taking $\delta = 6.0$, we plot the $T_c$ versus $J_s$ in different transverse field ${\it \Omega}/J_b=0.0$, ${\it \Omega}/J_b=2.0$ and ${\it \Omega}/J_b=4.5$. It is interesting that the curve of ${\it \Omega}/J_b =6.5$ overlap the axis of $J_s$ when $J_s$ is smaller than a certain value. The reason is that: the transverse field ${\it \Omega}$ suppress the orderness of the system. When transverse field is large enough, to a certain point, i.e. ${\it \Omega}_c$, the phase transition point tends to zero for small $J_s$, i.e. there is no ordered phase. While $J_s$ is larger than the value of $J_{sc}$, the system can be ordered then. In ferroelectric film, it is of importance that we should take the long-range coupling interaction into consideration. The long-range interaction makes the transition temperature higher.

\vskip 0.5cm
{\bf Acknowledgment}

Project was supported by the National Natural Science Foundation of China (Grant No.10174049) and the Natural Science Foundation of Jiangsu Education Committee of China (Grant No.00KJB140009).

\newpage
\begin{center}{Captions of Figures}\end{center}
{\bf Fig. 1.} The schematic of a multilayer ferroelectric film
\\
\\
{\bf Fig. 2.} The temperature as a function of the film thickness n in the spin-1/2 Ising model with transverse fields ${\it \Omega}/J_b=0.0$ (solid line) and ${\it \Omega}/J_b=2.0$ (dashed line)
\\
\\
{\bf Fig. 3.} The phase diagram ($T_c$ versus ${\it \Omega}$ plot) of the spin-1/2 10-layer system with a transverse field ${\it \Omega}$
\\
\\
{\bf Fig. 4.} The dependence of $T_c$ on the surface interaction constant $J_s$. (a) for three selected $\delta=3.2,6.0,1000$ with zero transverse field, (b) for three selected transverse fields: ${\it \Omega}/J_b=0.0, {\it \Omega}/J_b=3.0, {\it \Omega}/J_b=6.5$ with decaying exponent $\delta=6.0$\\

\newpage
\vfil\includegraphics[scale=0.7]{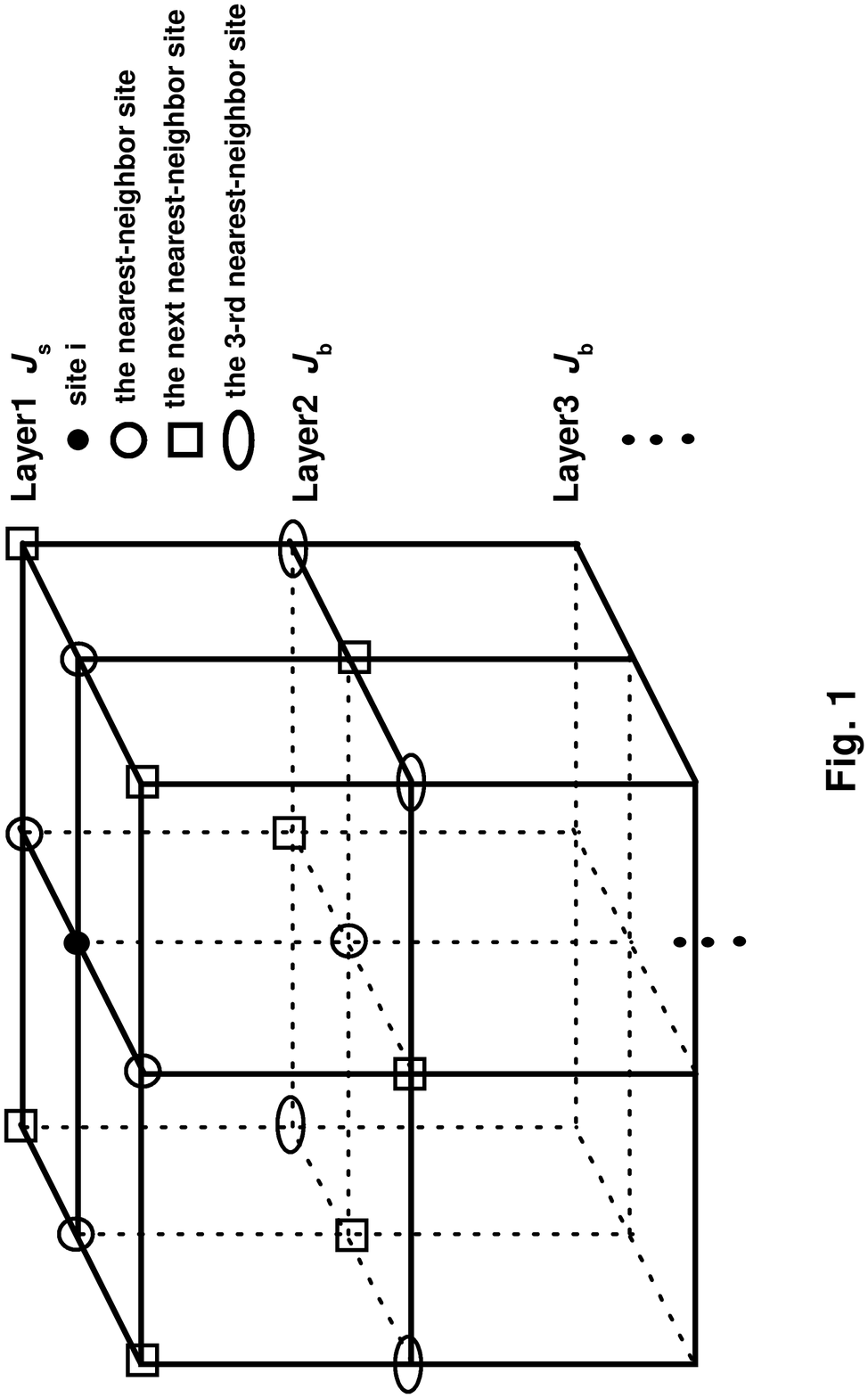}\vfil

\newpage
\vfil\includegraphics[scale=0.7]{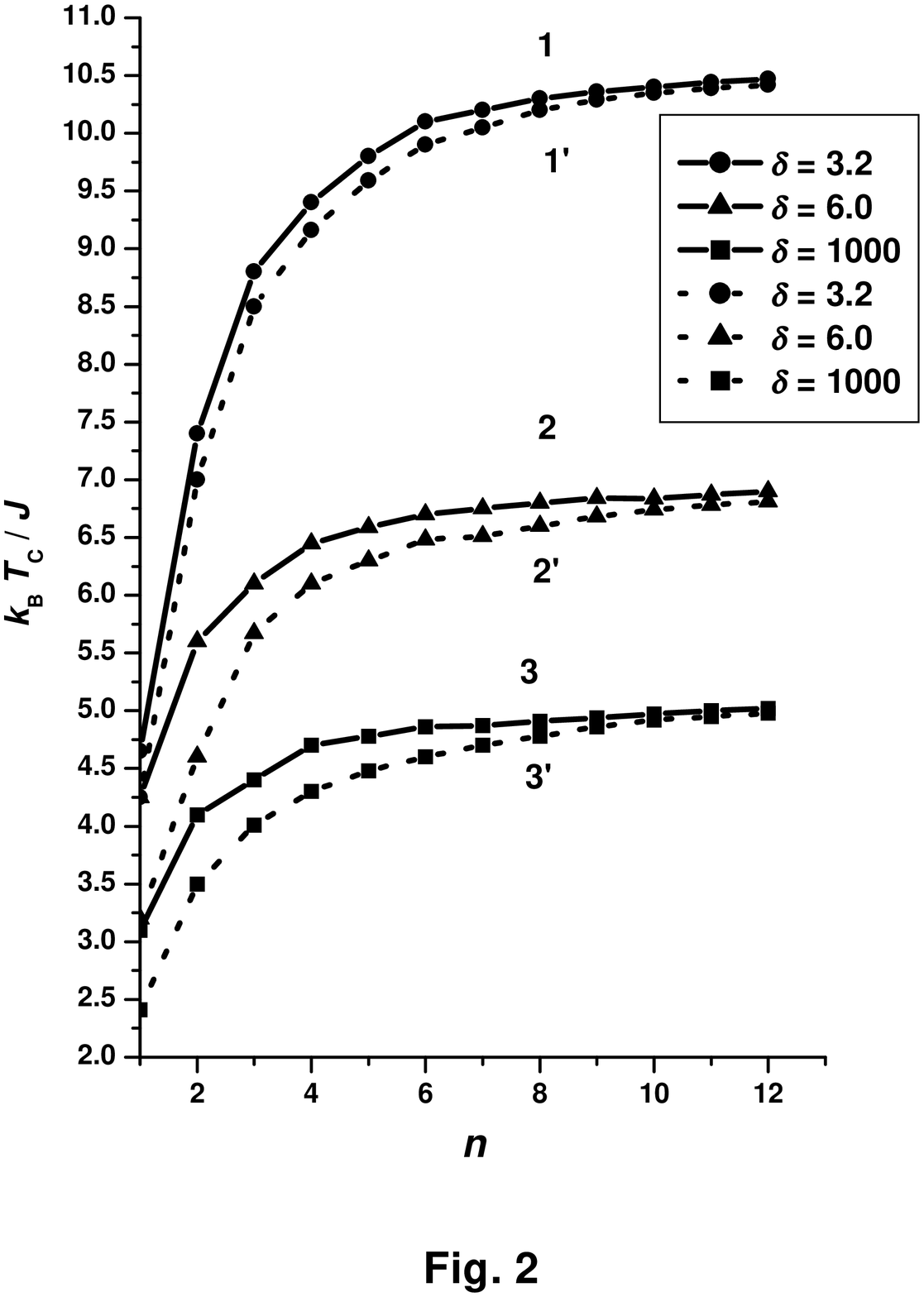}\vfil

\newpage
\vfil\includegraphics[scale=0.7]{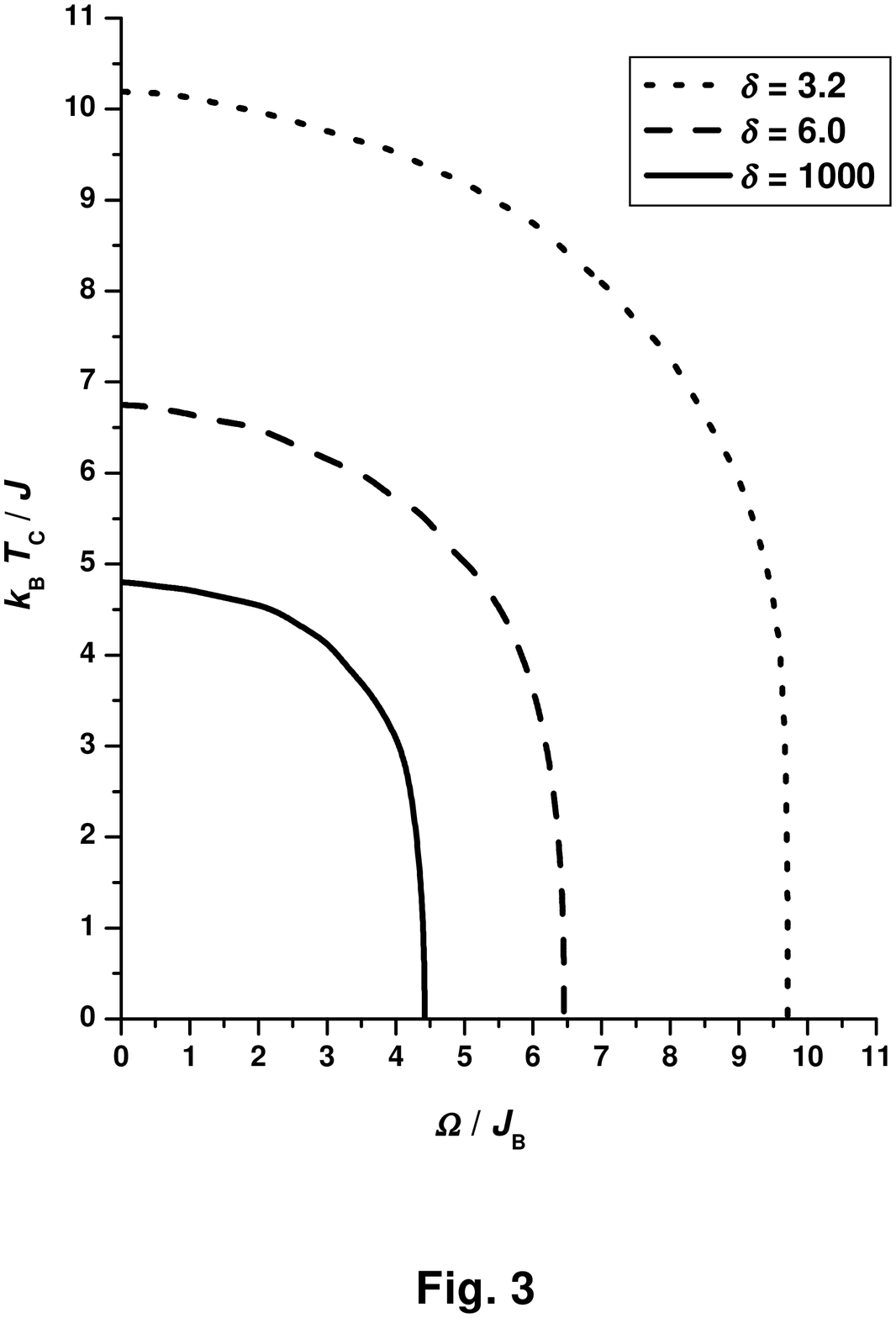}\vfil

\newpage
\vfil\includegraphics[scale=0.7]{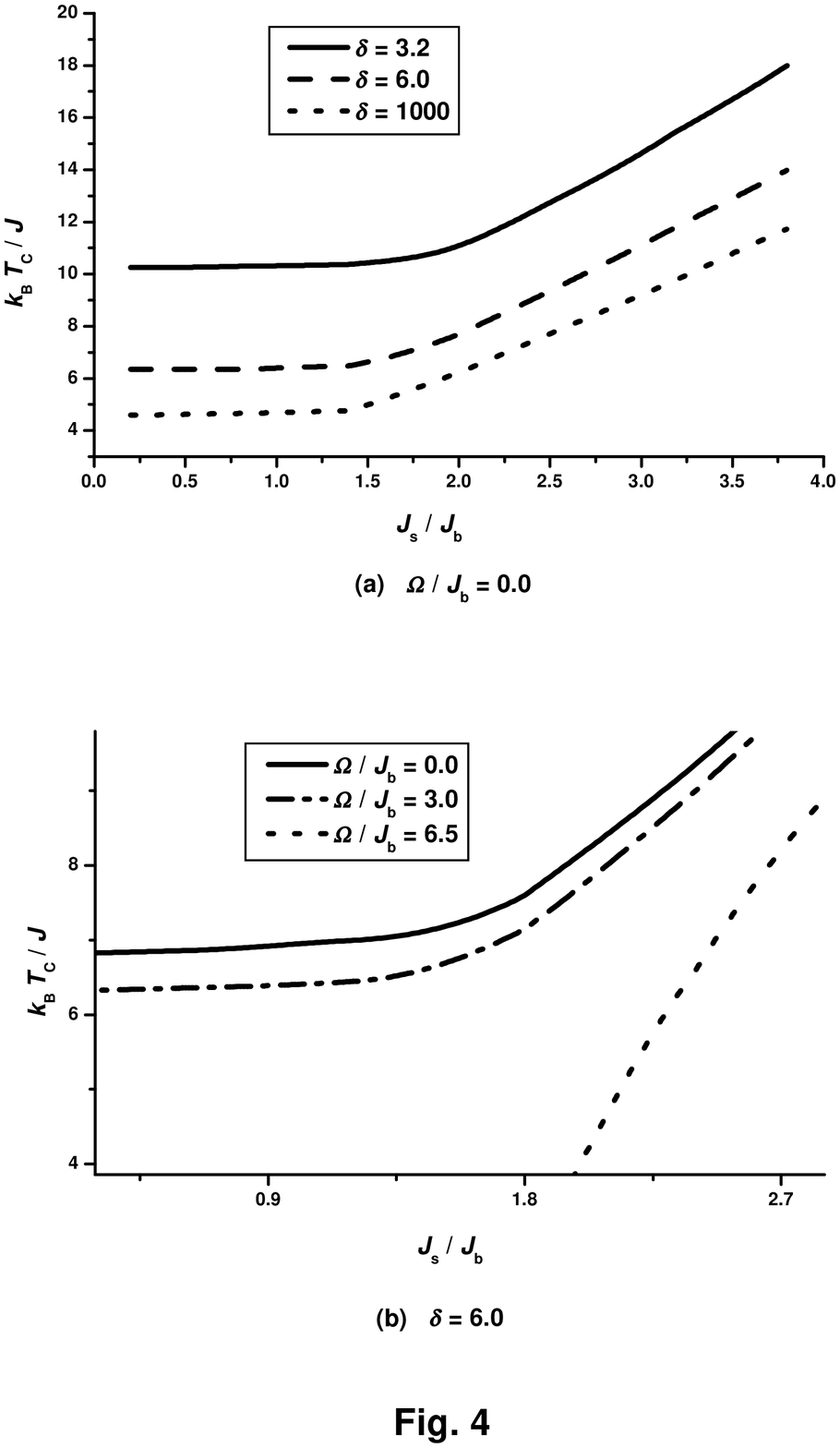}\vfil

\begin{thebibliography}{99}

\bibitem{}{\sc P. G. Gennes}, Solid State Commun. {\bf 1}, 132 (1963).

\bibitem{}{\sc R. Blinc} and {\sc B. Zeks}, Soft Modes in Ferroelectrics and Antiferroelectrics, North-Holland, Amsterdam 1974.

\bibitem{}{\sc R. B. Stinchcombe}, J. Phys. C: Solid State Phys. {\bf 6}, 2459 (1973).

\bibitem{}{\sc B. J. S. Lage} and {\sc R. B. Stinchcombe}, J. Phys. C: Solid State Phys. {\bf 12}, 1319 (1979).

\bibitem{}{\sc T. Kaneyoshi and S. Shin}, phys. stat. sol. (b) {\bf 218}, 537 (2000).

\bibitem{}{\sc T. Kaneyoshi}, phys. stat. sol. (b) {\bf 220}, 951 (2000).

\bibitem{}{\sc J. M. Wesselinowa}, phys. stat. sol. (b) {\bf 223}, 737 (2001).

\bibitem{}{\sc H. K. Sy}, J. Phys.: Condens. Matter. {\bf 5}, 1213 (1993).

\bibitem{}{\sc M. E. Lines} and {\sc A. M. Glass}, {\it Principles and Applications of Ferroelectrics and Related Materials}, pp.48. Clarendon, Oxford 1977.

\bibitem{}{\sc J. Shen and Y. Q. Ma}, J. Appl. Phys. {\bf 89}, 5031 (2001); Phys. Rev. B {\bf61}, 14299 (2000);\\{\sc Yin-Zhong Wu} and {\sc Zhen-Ya Li}, Chinese Physics {\bf 10}, 1058 (2001).

\bibitem{}{\sc Y. G. Wang}, {\sc W. L. Zhong} and {\sc P. L. Zhang}, Solid State Commun. {\bf 101}, 807 (1997).

\bibitem{}{\sc T. Balcerzak} and {\sc T. Kaneyoshi}, Physica A {\bf 206}, 176 (1994).

\bibitem{}{\sc T. Kaneyoshi} and {\sc H. Beyer}, J. Phys. Soc. Jpn. {\bf 49}(1980)1306;\\
{\sc T. Kaneyoshi}, J. Phys.: Condensed Matter {\bf 11}, 7311 (1999).

\bibitem{}{\sc X. Z. Wang} and {\sc Y. Zhao}, Physica A {\bf 193}, 133 (1993);\\
{\sc S. C. Lii} and {\sc X. Z. Wang}, Phys. Rev. B {\bf 51}, 6715 (1995); Physica A {\bf 232}, 315 (1996).

\bibitem{}{\sc J. H. Zhou} and {\sc C. Z. Yang}, Solid State Commun. {\bf 101}, 639 (1997).

\bibitem{}{\sc S. A. Cannas}, Phys. Rev. B {\bf 52}, 3034 (1995).

\bibitem{}{\sc M. E. Fisher}, {\sc S. K. Ma}, and {\sc B. G. Nickel}, Phys. Rev. Lett., {\bf 29}, 917 (1972).

\bibitem{}{\sc T. Kaneyoshi}, Acta Phys. Polon. A {\bf 83}, 703 (1993).

\bibitem{}{\sc F. C. SaBarreto}, {\sc I. P. Fittipaldi} and {\sc B. Zeks}, Ferroelectrics {\bf 39}, 1103 (1981).
\end{thebibliography}
\end{document}